\documentstyle[preprint,aps,prl,multicol,epsf]{revtex}
\tightenlines
\begin{document}
\draft
\title{Scale Dependent Dimension of Luminous Matter in the Universe}
\author{Per Bak{*}{$\dagger$} and Kan Chen{*}{$\dagger$}} 
\address{{$\dagger$}Niels Bohr Institute, Blegdamsvej 17, Copenhagen, 
Denmark.\\
{*}Department of Computational Science, Faculty of Science, National
University of Singapore, Singapore 117543
\\}
\date{\today}
\maketitle
\vspace{.1truein}

\begin{abstract} 

We present a geometrical model of the distribution of luminous matter in the 
universe, derived from a very simple reaction-diffusion model of turbulent 
phenomena. The apparent dimension of luminous matter,
$D(l)$, depends linearly on the logarithm of the scale $l$ under which
the universe is viewed: $D(l) \sim 3\log(l/l_0)/\log(\xi/l_0)$, where $\xi$ is
a correlation length. Comparison with data from the SARS red-shift catalogue, 
and the LEDA database provides a good fit with a correlation length 
$\xi \sim 300$ Mpc. The geometrical
interpretation is clear: At small distances, the 
universe is zero-dimensional and point-like. At distances of 
the order of 1 Mpc the dimension is unity, indicating a filamentary, 
string-like structure; when viewed at larger scales it gradually becomes  
2-dimensional wall-like, and finally, at and beyond the correlation length, 
it becomes uniform.
\end{abstract}

Uniformity of the 
background radiation requires that the universe must be homogeneous at the 
largest scale; this is known as the cosmological principle. However,
a decade ago, Coleman and Pietronero \cite{Coleman} suggested that the 
universe, at length scales $L$ up to a couple of Mpc is fractal with  
fractal dimension, $D\sim 1.2$, based on a study of the CfA galaxy
catalogue. Subsequent studies seemed to confirm this picture: Guzzo et al.
\cite{Guzzo} found $D = 1.2$ for  $L = 1-3$ Mpc, increasing to $D=2.2$  for 
$L = 3-10$ Mpc, from the Perseus-Pisces catalogue. Martinez and Coles 
\cite{MC} found that the dimension gradually increases from 2.25 to 2.77 at 
length scales increasing from $1 - 50$ Mpc. These empirical studies have  
recently been reviewed by Wu et al \cite{Wu}. Even though there is a general
agreement about the existence of fractal galactic structures at moderate
scales, there is still intense debate whether or not the universe is 
homogeneous at very large scales and, if so, how the transition to homogeneity
takes place \cite{Gabrielli}\cite{Gaite}. The value of the homogeneity scale 
and the matter distribution within such scale have great cosmological 
consequences.

We propose that the distribution of luminous matter in the 
Universe can be described by a new geometric scaling form that we discovered 
recently\cite{CB} in a different context. This description 
leads to a reconciliation of observational data at various scales and a 
consistent phenomenology of the crossover to homogeneity. A sharp transition
to homogeneity at $300$Mps is predicted.

The model is a simple non-equilibrium reaction-diffusion ``forest-fire'' model 
\cite{BCT}\cite{ff}, proposed to capture the essential features of turbulent 
systems, where energy is injected at the largest scale, and dissipated at a 
small length scale. 
We found that in a range of length scales between these 
two limits the dimension of the luminous field (fire distribution) varies 
gradually from  zero to three, and that the distribution becomes homogeneous
beyond a correlation length which depends on the energy injection rate. 
The model operates near a dynamical ``critical point'', with diverging 
correlation length. 

As we will show below, analysis of galaxy maps indicates that the geometrical 
structure of luminous matter in the universe is very similar to that of the 
forest-fire model. Our alternative form provides a better fit to
the data than conventional models. The underlying picture is one where
luminous matter is being created and destroyed in an ongoing 
non-equilibrium dynamical process. This similarity is appealing in that it 
suggests that the universe shares the basic characteristic
features of other dynamical systems, so perhaps the dynamics of the universe
is not unique, but belongs to a more general universality class of
non-equilibrium turbulent systems. 

Usually, systems near equilibrium criticality are self-similar, or fractal, 
for length scales below the correlation length; hence fractal behavior can 
often be 
viewed as a consequence of criticality \cite{BTW}. However, 
the forest-fire model does not show simple power-law (fractal) scaling 
below the correlation length. Numerical studies show that
the average amount of dissipation $n(l)$, 
within a cube box of size $l$ that contains dissipation, obeys 

\begin{equation}
\log(n) \sim \Bigg({3\over 2}{\log(l/l_0)\over \log(\xi/l_0)}\Bigg)\log(l/l_0) 
\quad ,
\end{equation}
where $l_0 \sim 1$ lattice spacing for the forest fire model. At the 
correlation length $\xi$, there is a sharp cross-over to a homogeneous 3d 
structure. Thus, the correlation length is identical to the homogeneity 
length, as is
usually the case in critical phenomena. However, the lack of self-similarity 
implies that one can derive the correlation length from observations in a 
range of much smaller length scales.

This equation can be interpreted in terms of an  apparent fractal dimension 
for luminous matter that varies linearly with the logarithm of the length 
scale:
\begin{equation}
D(l)={d\log(n)\over d\log(l)} \sim 
3\Bigg({\log(l/l_0)\over \log(\xi/l_0)}\Bigg) \quad.
\end{equation}

We suggest that the length scale
dependent behaviour observed in this model may be sufficiently general that it
is worthwhile to make a detailed comparison with real astronomical data. The
underlying viewpoint is that the galactic dynamics is turbulent,
with stellar objects interacting with one another in reaction-diffusion type 
processes through shock waves, super-novae explosions, galaxy mergers, etc.
Apart from an overall amplitude, there are only two fitting parameters in our
proposed galaxy distribution, namely the upper
length scale $\xi$, where the distribution becomes uniform, and the lower 
cutoff, $l_0$, where the distribution becomes point like.

In their seminal work, Sylos-Labini, Pietronero and coworkers \cite{Pietronero}
analysed several database catalogues of galaxy maps. From the databases, they 
created volume-limited (VL) samples containing all galaxies exceeding a 
certain absolute luminosity within a given volume. Then they calculated the 
conditional density  $\Gamma^*(l)$, which is the average density of galaxies 
within a sphere of size $l$. This quantity corresponds to the density $n(l)$
defined above, divided by the volume $l^3$. Thus, the resulting prediction for
$\Gamma^*(l)$ becomes

\begin{equation}
\log\big(\Gamma^*(l)\big) \sim 
\Bigg({3\over 2}\Big({\log(l/l_0)\over \log(\xi/l_0)}\Big)-3\Bigg)\log(l/l_0) 
\quad .
\end{equation}

Namely, on a log-log plot there is a pure quadratic dependence, rather than 
the linear dependence found for self-similar fractal structures.

We have fitted the above expression to the conditional densities extracted
by Pietronero et al. from two widely different data bases with consistent 
results. The LEDA database is a heterogeneous compilation of data from the 
literature containing more than 200,000 galaxies. The Stromlo-APM red shift 
survey (SARS \cite{SARS}) consists of 1797 galaxies. Figure 1 shows results from the fits,
with two different cut-offs for the LEDA database.
The labeling follows Sylos Labini et al., with the numbers representing the 
lower luminosity cut-offs.
Obviously, there are larger fluctuation for the sparser, 
but perhaps higher quality, Stromlo-APM data set than for the LEDA database. 
The fits are very good in view of the fact that the only fitting 
parameters are the upper and lower length scales, $\xi$ and $l_0$, 
respectively. In contrast to conventional critical phenomena, the correlation
length enters the expression for length scales below the correlation length. We
are therefore able to fit the correlation length to the data, despite the
fact that no data is available at and beyond the projected correlation
length.

The upper length scale is the one where the curves become flat, $d=3$. The 
three fits yield very consistent values of this length scale, $\xi = 260$ Mpc 
from the LEDA16 data, $\xi = 275$ Mpc from the LEDA14 data, and $\xi = 380$ 
Mpc from the APM data.

The empirical logarithmic scale dependence of the dimension can be seen 
directly by re-plotting the data in figure 1:
Figure 2 shows $D(l)=2 \times (\log (\Gamma^*(l)/\Gamma(l_0))/\log(l/l_0) +3)$.
All data sets yield linear behavior. The correlation length is 
found by linear extrapolation to the point where $D(l)$ assumes the value of 
3. The dimensions derived from the intense galaxies, LEDA 16 and APM 18, are 
essentially identical, but the LEDA 14 data yield a
somewhat steeper scale dependence. However, they all converge at essentially 
the same homogeneity length.

We predict a sharp crossover to uniformity, i. e. a sharp kink in the curve, 
at this length, which  will be readily observable once data becomes available.
Actually, 
there is a recent analysis 
based on ESO Slice Project galaxy redshift survey which indicates that the 
fractal 
dimension is close to 3 for the length scale greater than
300 Mpc \cite{Scaramella}). Also, the intermediate data points are predicted
to follow the straight lines in figure 2. 

The lower cut-off, $l_0$, is the scale at which the slope of the curves in 
the figure assumes the value of -3. We find $l_0=370$ light-years, $l_0=3700$
light-years, and $l_0=330$ light years for the three samples, respectively.
This scale is determined with less precision than the correlation length $\xi$.
It is not clear how well our scaling form applies to the analysis of the
galaxy distribution at small length scales.

The geometry of the luminous set is {\it not fractal} when viewed over the 
entire range of scales, since there is no 
self-similarity for different scales. Nevertheless, the scale dependent 
dimension has a clear geometrical interpretation: At small distances, the 
universe is zero-dimensional and point-like. Indeed, energy dissipation takes 
place on individual point objects, like stars and galaxies. At distances of 
the order of 1 Mpc the dimension is unity, indicating a filamentary, 
string-like structure; when viewed at larger scales it gradually becomes  
2-dimensional wall-like, and finally at the correlation length, $\xi$, it 
becomes uniform.

It might be instructive to compare with more conventional interpretations of
the large scale structure \cite{Peebles}. The conditional density can be 
related to a correlation function $g(r)$ through \cite{Pietronero} 

\begin{equation}
\Gamma^*(l) \sim <n>(1+g(l)),
\end{equation}  

where $<n>$ is the mean density of galaxies. For instance, the field theory
of de Vega et al. \cite{Sanchez} yields an expression of this form. The correlation
function is often assumed to be of the form
$g(l) = (r_0/l)^{\gamma}$. Figure 1 also shows a fit to this expression,
with $r_0 = 10$ Mpc and $\gamma = 1.3$. The fit is clearly inferior,
flattening out at too small length scales. This is in accordance with the 
observations by Sylos Labini et al. that the value of fitted parameter $r_0$
depends heavily on the range of length scales used. At larger scales, the 
difference between the two fits is even more pronounced; when further data 
becomes available in the near future, one should be able to discriminate even 
better between the two pictures. In this traditional view, there is a 
smooth crossover to homogeneity when the {\it amplitude}, expressed in terms 
of $r_0$ reaches unity. In contrast, following our ``critical phenomena'' 
viewpoint, there is a sharp, possibly exponential, cutoff of the non-uniform 
part of the correlation function at the {\it correlation length}.

This has some important cosmological consequences. In the traditional 
formulation, one usually visualizes that the amplitude $r_0$ of the power-law 
fluctuations increases with time, starting from the time of the decoupling
of radiation from hadronic matter, leading to an increase of the cross-over  
length to homogeneity. In our phenomenology, the correlation 
length $\xi$ is the only parameter, so it is this quantity which is increasing 
with time. The average density of galaxies in the universe is equal to the 
density within the correlation length, i.e. $<n> = \Gamma^*(\xi) \sim 1/\xi^{3/2}$. 
Thus, once the correlation length has been determined, one knows the density
of galaxies. Assuming that the entire density of hadronic matter scales as that
of the luminous galaxies studied here, one might get an estimate of the mass 
of the universe.  From the fit to LEDA 14 one gets that the density of galaxies 
in the entire universe with apparent magnitude greater than 14 is 
$<n> = 2 \times 10^{-3} Mpc^{-3}$. From the fit to 
the APS 18 data we find that the density of galaxies with apparent luminosity 
greater than 18 is  $<n> = 3 \times 10^{-4} Mpc^{-3}$. The traditional fits give much  
larger values for the density of galaxies in the universe, depending on the 
range of length scales used in the fit \cite{Pietronero}.

In the forest fire model the energy flux (which determines the average 
density of fires) is an independent parameter, namely 
the growth rate of trees, whereas for the universe it is self-consistently 
determined by the dynamics. The forest fire model exhibits
self-organised criticality \cite{BTW}, in the sense that the correlation  
length diverges as the tree growth rate $p \rightarrow 0$. 
All fire goes extinct as the correlation length reaches the system size.
As the universe expands, the correlation length $\xi(t)$ 
increases faster than the size of the universe $R(t)$
and the universe become more and more inhomogeneous.
One might speculate that as the correlation length reaches the size of the 
universe, all the luminous matter is extinguished, and we are left with a 
universe without luminous matter!
 
{

\pagebreak

{\bf Acknowledgment}. We thank Maya Paczuski, Kim Sneppen, and Jakob Bak for 
helpful discussions and comments on the manuscript.
\vspace{20mm}

{\bf Figure Captions}
 
\vspace{6mm}

\noindent Figure 1. Conditional average densities for various galaxy catalogues 
(arbitrary scale),
as derived by Sylos Labini et al \cite{Pietronero}, compared with fits to 
equation 1, yielding $\xi = 260$ Mpc from the LEDA16 data, $\xi = 275$ Mpc 
from the LEDA14 data, and $\xi = 380$ Mpc from the APM data. The broken line
is a conventional fit to equation 4 with $\gamma = 1.3$, $r_0 = 10$ Mpc.

\vspace{6mm}

\noindent Figure 2. Scale dependent dimension $D(l)$ derived from the
data points in figure 1 as explained in text. We conjecture that future data
points follow the straight lines, and saturate sharply to $D=3$ at the
correlation length.


\begin{references}
\bibitem{ff}
The model is  a ``forest fire model'' defined as follows. On a 3-dimensional 
cubic lattice, trees, representing energy, are grown randomly at empty sites 
at a rate $p$. During one time unit, trees burn down (leaving room for new 
trees), and ignite neighbouring trees. These processes  represent dissipation 
and diffusion of energy, respectively. After a transient period, the system enters a statistically stationary state with a complex distribution of fires. It
is this distribution that we compare with the distribution of luminous matter.
\end{references}
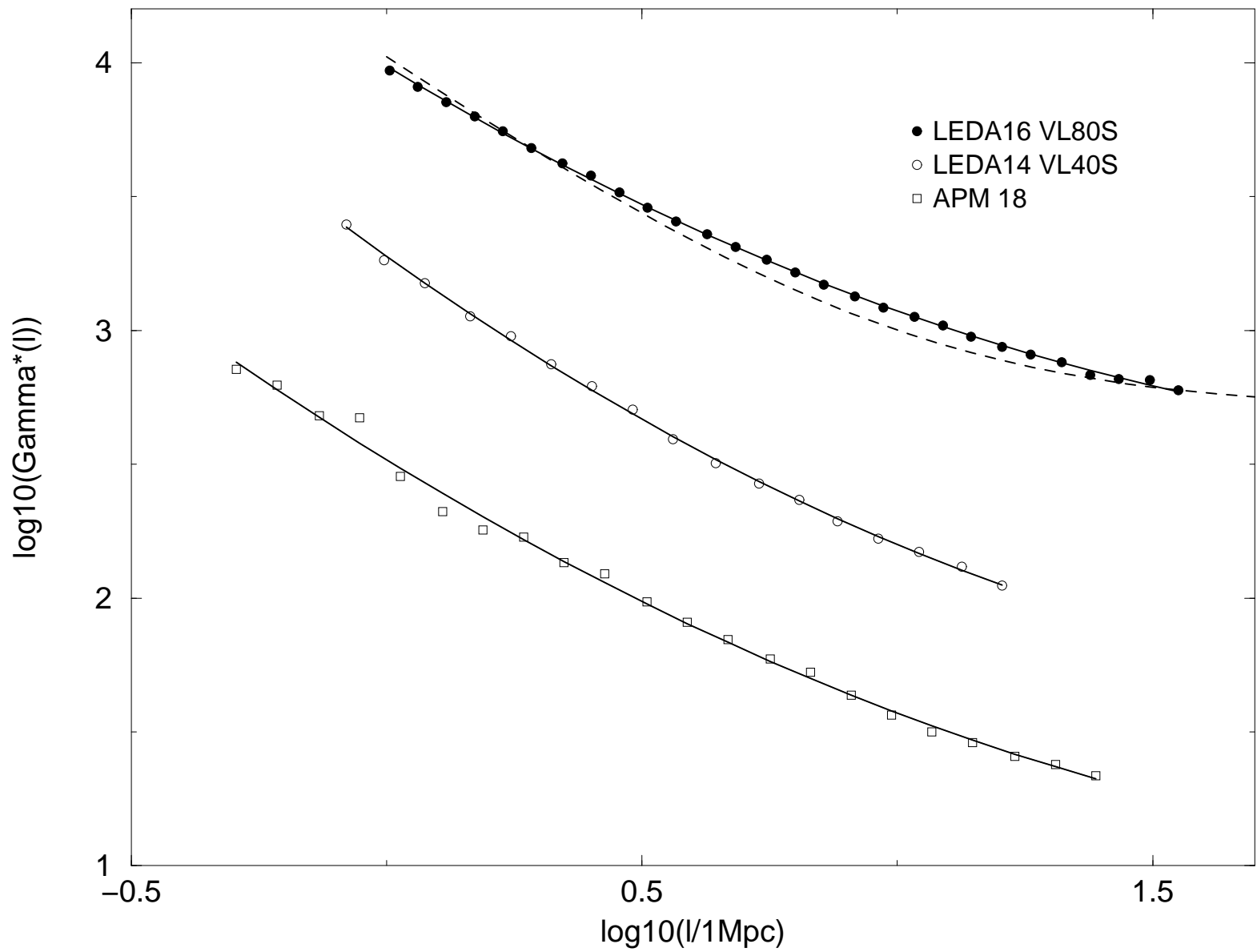
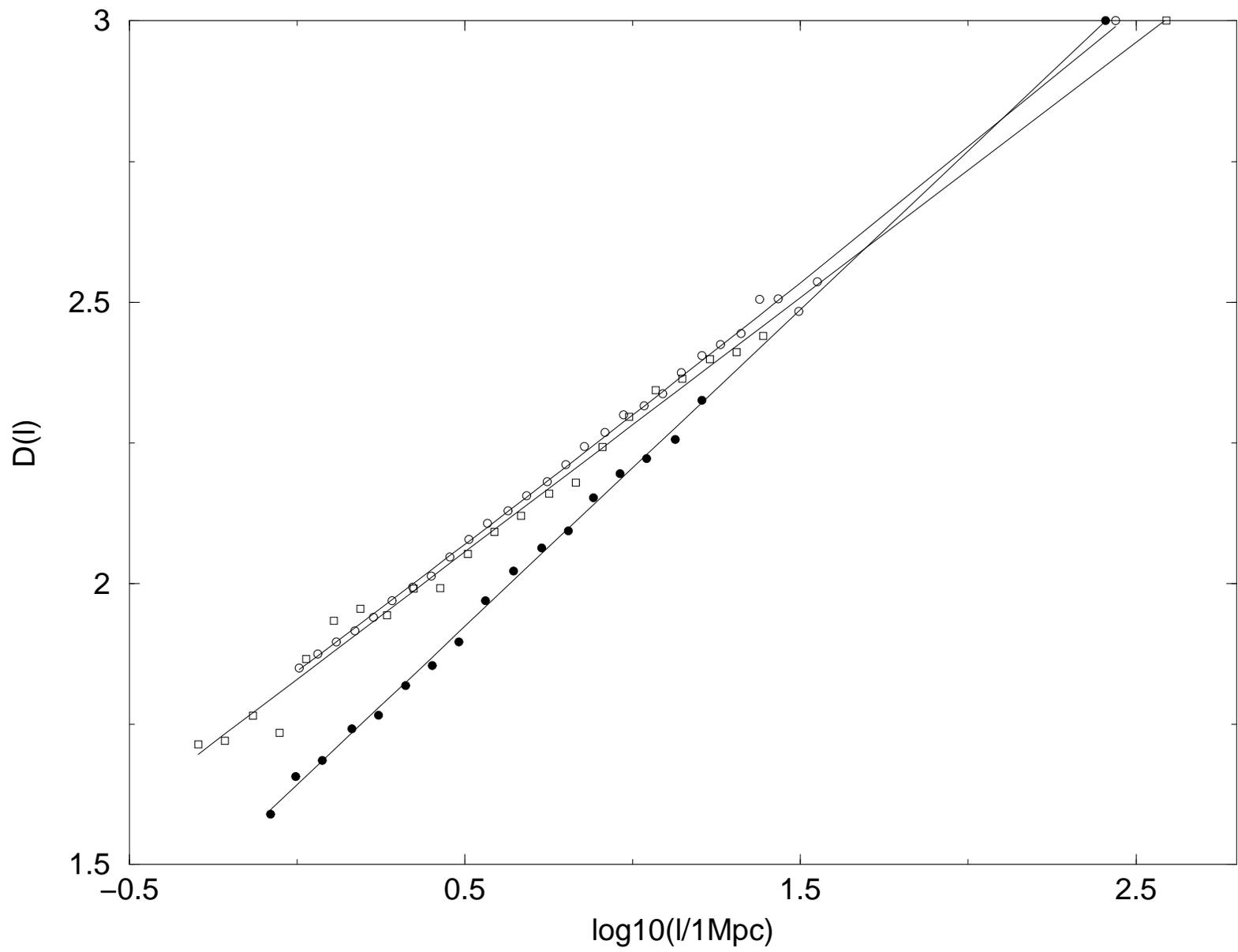
\end{document}